\parskip=0pt

\def\R{{\bf R}}
\def \d{{\rm d}}

\def \ep{\epsilon}
\def \( {\big( }
\def \) {\big) }
\def \gs {\lower3pt \hbox {${\buildrel > \over \sim}$}}

\def \nhy {nonhyperbolic}

\def \q {\quad}
\def \qq {\qquad}
\def \pd{\partial}
\def \pn{\par\noindent}
\def \bs{\bigskip}
\def \en{\eqno}
\def \Ref{\pn{\bf References}\bs\pn\parskip=5 pt\parindent=0 pt}
\baselineskip 0.55 cm

\def\.#1{\dot #1}

{\nopagenumbers
~ \vskip 2 truecm
{\bf \centerline {Nonhyperbolic homoclinic chaos}}
\vskip 1.5 truecm
\centerline {{\bf G. Cicogna}\footnote{$^{(*)}$}{E-mail : 
cicogna@difi.unipi.it} and {\bf M. Santoprete}}
\centerline{Dipartimento di Fisica, Universit\`a di Pisa, }
\centerline{Via Buonarroti 2, Ed. B}
\centerline{I-56127, Pisa, Italy}

\vskip 2 truecm

\pn
{\bf Abstract.} 
\pn
Homoclinic chaos is usually examined with the hypothesis of hyperbolicity 
of the critical point. We consider here, following a (suitably adjusted) 
classical analytic method, the case of non-hyperbolic points
and show that, under a Melnikov-type condition plus an additional 
assumption, the negatively and positively asymptotic sets persist 
under periodic perturbations, together with their infinitely many 
intersections on the Poincar\'e section. 
We also examine, by means of essentially the same procedure, the case of 
(heteroclinic) orbits tending to the infinity; this case includes in 
particular the classical Sitnikov 3--body problem.

\bs\bs\pn
PACS. no: 03.20, 05.45 
\pn
{\it Keywords}: homoclinic chaos, nonhyperbolic critical point, Melnikov 
theory, Sitnikov problem.
\vskip 2 truecm\pn
{\hfill To appear in Phys. Letters A}

\vfill\eject }
\pageno=2

~ \vskip 3 truecm

It is well known that perturbing homoclinic (or heteroclinic) orbits of 
dynamical systems may lead to the phenomenon of transversal 
intersections of stable and unstable manifolds of the 
critical point, and that this is one of the routes for the 
appearance of chaos; it is also known that Melnikov method is a very
efficient analytical criterion to determine the occurrence of the intersection 
of stable and unstable manifolds. It is impossible to give here even a partial 
account of the vast literature on this subject, and we will 
quote here only those papers [1-7] which are more strictly related to the 
present approach (for a more complete list of references, see also [8]).

It can be noticed, however, that almost the totality of the papers 
dealing with homoclinic chaos and Melnikov method considers only the case 
of {\it hyperbolic} critical points (some exceptions will be quoted 
below). We want to show here that a suitable extension of a classical and 
purely analytical method, used in [5] for the  hyperbolic case, can also 
cover -- under suitable hypotheses -- the \nhy\ case.

For definiteness, even if more general situations could be considered, 
we will deal here only with planar systems originated from unperturbed 
Hamiltonians  of the form
$$H={1\over 2} p^2 + V(x)\en (1)$$
We point out that this method can (and will) be applied not only to the 
case of homoclinic orbits approaching \nhy\ points corresponding to some 
``superquadratic'' unstable equilibrium points $x_0\in\R$ of the 
unperturbed potential $V(x)$, but also to the quite different case of
(heteroclinic) orbits tending to $\pm\infty$ for 
$t\to\pm\infty$ respectively, appearing in the case of potentials having 
an equilibrium point at the infinity.

For the sake of concreteness, we shall provide the construction and the 
proof only for the case of superquadratic stationary points, but 
it will appear clear that the argument can be repeated equally well for 
the other one. For what concerns the first case, we refer also to [9], 
where however a completely different approach is used; the second situation 
includes in particular the classical Sitnikov restricted $3-$body problem 
in celestial mechanics, which has been considered in detail by Moser [4], 
by means of the introduction of a singular change of coordinates 
(see also [10,11]).

More specifically, we will show that the negatively 
and positively asymptotic sets of the unstable equilibrium point persist 
even in the \nhy\ case and under periodic perturbation. We also obtain that 
the occurrence of an intersection of these sets on the Poincar\'e section 
(together with the chain of their infinitely many subsequent intersections)
can be detected by a Melnikov criterion, plus an additional condition, 
which must be introduced in this case to compensate the lack of the
``exponential dichotomy'' peculiar of the hyperbolic case.

\vfill\eject\pn
{\bf 1.} We consider a Hamiltonian dynamical problem with Hamiltonian 
of the form (1), plus a smooth perturbation $g$ depending on one or more 
real (small) parameters $\ep$:
$$\eqalign { \.x= & \ p \cr \.p=& -{\d V\over {\d x}} + g(\ep,x,\.x,t)
\qq {\rm with}\q g(0,x,\.x,t)=0 } \en(2)$$
where the potential $V(x)$ of the unperturbed problem is assumed to be 
analytical and to have a ``superquadratic'' unstable equilibrium point at 
$x=x_0$ (say, $x_0=0$), which may be either a ``cubic-like'' 
stationary point, or a local maximum for $V(x)$ (in this case, clearly,
$m<0$ in (3) below):  
$$V(0)={\d V(0)\over{\d x}}=\ldots{\d^{\nu-1} V(0)\over{\d x^{\nu-1}}}=0 
\q ; \q {\d^{\nu} V(0)\over{\d x^{\nu}}}=m\not=0 \q {\rm for\ some \ integer}
\ \nu>2 \en(3)$$
We assume that the unperturbed problem possesses a homoclinic orbit 
doubly asymptotic to $x_0=0$:
$$\chi=\chi(t) \qq {\rm with} \qq \lim_{t\to\pm\infty}\chi(t)=0\en(4)$$
this happens, e.g., if $V(x)$ also admits at least a stable equilibrium point 
$x_1$ (e.g., with $x_1>x_0$), and there is another point $x_2>x_1$ such 
that $V(x_2)=0$ and $\d V(x)/\d x<0$ for $x_0<x<x_1$, $\d V(x)/\d x>0$ for 
$x_1<x\le x_2$.

A typical  example of this situation is given by a double-well
Duffing-type potential, modelling a great number of physical situations
(this example will be considered explicitly at the end of this paragraph)
$$V(x)=-{1\over 2}x^4+{1\over 2}x^6\en(5)$$
It is immediate to see that the unstable equilibrium point $x_0=0$ \big(or 
better, the point $x=0,\ p=0$ of the first-order problem (2), with a 
potential satisfying (3)\big) is not a hyperbolic 
point, indeed the linearized part of (2) (with $\ep=0$) is given by 
the nilpotent matrix 
$$\pmatrix{ 0 & 1 \cr 0 & 0 } $$
(cf. [9,12], where existence and regularity of the invariant manifolds 
is investigated in a very general setting, see also [11]).
A first consequence of this fact is that now the ``exponential dichotomy'' 
peculiar of the hyperbolic situation is no longer available; 
it is easy to see in fact that the homoclinic solution of the unperturbed 
problem vanishes in our case for $|t|\to\infty$ as
$$\chi(t) \sim |t|^{-2/(\nu-2)}\en(6)$$
Consider now the effect of the perturbation $g(\ep,x,\.x,t)$: here, we must 
assume first of all that
$$g(\ep,0,0,t)=0\en(7)$$
this ensures that  $x_0=0$ is an equilibrium point also for the 
perturbed problem. The non-hyperbolicity of this point cannot ensure a 
priori the existence of stable and unstable manifolds; more precisely, 
the non-hyperbolicity has a more radical consequence: we cannot 
even properly speak of stable and unstable manifolds, indeed, the 
eigenvalues of the linearized problem are zero, and all the dynamics 
occurs actually on the {\it center} manifold (cf. [13]).

We now want to show that  suitable assumptions on the perturbation $g$ may 
guarantee not only the existence of solutions  approaching the 
equilibrium point $x_0=0$ for $t\to -\infty$ and  
for $t\to +\infty$, which we will denote respectively by 
$$x_o=x_o(t) \qq {\rm and} \qq x_i=x_i(t)$$  
forming the ``out-set'' and the ``in-set'' -- and playing the role of 
negatively and positively asymptotic sets of the hyperbolic case --
but also the possible presence of transversal intersections on the 
Poincar\'e sections.

Let us write the problem (2) in the form (where $f=-\d V/\d x$)
$$\ddot x-f(x)=g(\ep,x,\.x,t)\en(8)$$
and let us look for solutions $x(t)$ of the perturbed problem ``near'' 
the family of the homoclinic orbits $\chi(t-t_0)$: we then put for convenience
(see [5])
$$x(t)=\chi(t-t_0)+z(t-t_0) \en(9)$$
Inserting into (8) (with the time shift $t-t_0\to t$), we obtain
$$\ddot z-f'[\chi(t)]z(t)=G(\ep,z,\.z,t+t_0)\en(10)$$
where the r.h.s. is given by (we will shortly denote by $G=G(\ep,t,t_0)$ this 
quantity; notice that no approximation has been made, nor any term neglected)
$$G=G(\ep,t,t_0)\equiv g\( \ep,\chi(t)+z,\.\chi(t)+\.z,t+t_0\) +
f\( \chi(t)+z\) -f\( \chi(t)\) -f'[\chi(t)]z \en(11)$$ 
Consider the homogeneous equation obtained from (10): one solution of  is 
clearly $\.\chi(t)$, another independent solution $\psi(t)$ can be 
constructed with standard methods (see e.g. [14]) or by direct substitution: 
these two solutions have a different behaviour for $t\to\pm\infty$,
precisely:
$$\eqalign { \.\chi(t) \sim & \ |t|^{-\nu/(\nu-2)}\ \to 0  \cr
           \psi(t) \sim & \ |t|^{2(\nu-1)/(\nu-2)}\to\infty}\en(12)$$ 
The general solution $z(t)$ \big(even if in implicit form, through the 
definition (11) of $G$\big) of the complete equation (10) can be written
in the following integral form, 
with $A,B$ arbitrary constants and $t_1$ arbitrarily fixed:
$$z(t)=A\.\chi(t)+B\psi(t)-\.\chi(t)\int_{t_1}^t \psi(s)G(\ep,s,t_0)\ \d s +
\psi(t)\int_{t_1}^t \.\chi(s)G(\ep,s,t_0)\ \d s \en (13)$$
Let us now look for solutions $z_o(t)$ \big(and resp. $z_i(t)$\big) of (13)
with the property of vanishing for $t\to-\infty$ (resp. $t\to+\infty$); 
these will provide solutions 
$$x_{o,i}=\chi(t)+z_{o,i }(t)$$ 
of (8) belonging, by definition, to the  out-set  (resp. the in-set) of 
$x_0=0$. We have then to require that
$$\eqalign{
\lim_{t\to\mp\infty}\Big[\psi(t)\Big(B+\int_{t_1}^t \.\chi(s)G(\ep,s,t_0)\ 
\d s\Big) \Big]= 0
\cr
\lim_{t\to\mp\infty}\Big[\.\chi(t)\Big(A-\int_{t_1}^t\psi(s)G(\ep,s,t_0)\ 
\d s\Big)\Big] = 0}\en (14)$$
where the limits are to be evaluated at $-\infty$ when looking for 
the $x_o$ solutions, 
and resp. at $+\infty$ for the $x_i$ solutions. Notice that, in contrast 
with the hyperbolic case, we have to request here the vanishing of 
these quantities (not only their boundedness, as in [5]).

Now, it is an easy exercise to check, taking also into account the 
behaviour (12) of 
the two fundamental solutions $\.\chi(t)$ and $\psi(t)$, that the above 
conditions (14) are  satisfied  simultaneously at $t=-\infty$ and $+\infty$
if the following (Melnikov-type condition) 
$$\int_{-\infty}^{+\infty} \.\chi(t)G(\ep,t,t_0)\ \d t =0 \en(15)$$
and the additional condition
$$\lim_{t\to\mp\infty} \.\chi(t)\int_{t_1}^t\psi(s)G(\ep,s,t_0)\ \d s=0
\en (16)$$
are satisfied. Consider now the linearization of the problem (13) 
around the solution $z(t)\equiv 0,\ \ep=0$; in this way, condition 
(15) becomes a computable Melnikov condition 
$$M(\ep,t_0)\equiv\int_{-\infty}^{+\infty}
\.\chi(t)\ g\( \ep,\chi(t),\.\chi(t),t+t_0\) \ \d t=0 \en(17)$$  
and the same is true for the other condition (16). Once these are 
satisfied, it is sufficient to apply the implicit-function 
theorem to be granted (see [5])  that there is a regular solution 
of (10), or equivalently of (13), close to the 
unperturbed solution $z\equiv 0$, i.e. a solution of (8) close to the 
homoclinics $\chi(t-t_0)$ and approaching $x_0=0$ for $t\to\pm\infty$.

Assume now also that
$${\pd M\over{\pd t_0}}\not= 0 \en(18)$$
the derivative being evaluated at the  point $\ep, t_0$ where (17) is 
satisfied (this is of course ``generically'' true, see [15] for a careful 
discussion on non-transversal crossings) 
and that (16) also is satisfied (this may be easily  fulfilled, 
see below for this point): there is then, thanks also to (18), a 
transversal intersection at $t=t_0$ on the Poincar\'e section of the out/in 
asymptotic sets of $x_0=0$. Using then standard arguments (not based on 
hyperbolicity), considering the case of perturbations $g$ which are 
periodic (or almost periodic)  functions of the time,
one immediately deduces that there is an infinite sequence of 
transversal intersections, leading to a situation similar to the usual 
chain of homoclinic intersections typical of the homoclinic chaos. 

More exactly, we have now to emphasize that, in order to get a more 
complete description of the dynamical behaviour in this case, we cannot 
directly refer to the classical Birkhoff-Smale theorem on the equivalence  
with the Smale horseshoe dynamics, because the proofs of this equivalence 
are intrinsically based on hyperbolicity properties [2-4,6,7]. 
Some short comment about this point will be given at the end of this paper. 
Here, we can qualitatively say that the 
presence of infinitely many intersections indicates at least the presence
of a quite complicated geometry in the behaviour of the dynamical flow, 
i.e. a sort of ``\nhy\ homoclinic chaos'' (cf.  [9-13]). 

We now give some conditions to ensure that (16) is satisfied. We assume 
here for instance that the perturbation $g$ can be written in the form
$$g(\ep,x,\.x,t)= \ep_1x^{n_1}\sin\omega t + \ep_2\.x^{n_2}\en(19)$$
(notice, incidentally, that if the term with $\ep_2$ describes a damping 
and $n_2$ is even, then instead of $\.x^{n_2}$ one must more correctly 
write $|\.x|\.x^{n_2-1}$, with $\ep_2>0$, of course), it is easy to check, 
recalling also (12), that if 
$$ n_1>\nu-2 \qq {\rm and}\qq n_2>{2(\nu-2)\over \nu} \en(20)$$
then condition (16) is satisfied.
\bs
Considering e.g. the following explicit example, with $V$ given by (5),
$$\ddot x=2x^3-3x^5+\ep_1x^3\sin\omega t+\ep_2\.x^3\en(21)$$
there are two homoclinic orbits of the unperturbed Hamiltonian, given by
$$\chi(t)={\pm 1\over{\sqrt{1+t^2}}}$$
conditions (20) are satisfied (here $\nu=4$) and the Melnikov condition 
(17) can be explicitly evaluated, giving transversal homoclinic intersections 
if
$$\Big|{\ep_1\over{\ep_2}}\Big|>{3\exp(\omega)\over{32\ \omega(1+\omega)}} 
\en(22)$$
A numerical integration of this problem, performed along the same lines 
as in [6, Chapt.2]  (see [8] for details), shows that, choosing e.g. 
$\omega=1,\ \ep_2=.05$, transversal intersections appear when 
$\ep_1\gs .006$, in quite good agreement with the value $\ep_1=.00637$ 
given by (22).
\bs
Summarizing, we can state the above results in following form:
\bs\pn
{\bf Theorem.} Consider a planar dynamical system of the form (2) where 
the potential $V(x)$ admits a superquadratic unstable equilibrium in some 
point, say $x_0=0$, producing then a \nhy\ unstable 
equilibrium point for the unperturbed system (given by $\ep=0$). 
Assume that the unperturbed system admits a homoclinic orbit $\chi(t)$ doubly 
asymptotic to $x_0$. Assume for simplicity the perturbation $g$ of the 
form (19) with the conditions (20) satisfied. If in addition Melnikov 
conditions (17) and (18) are satisfied, then the perturbed problem admits  
a negatively and positively asymptotic sets of the unstable equilibrium 
point, which admit an infinite sequence of homoclinic intersections on 
the Poincar\'e section, giving rise to a (\nhy\ homoclinic) chaotic 
dynamical flow.

\bs\bs\pn 
{\bf 2.} Instead of discussing some of the possible extensions of the 
above results to more general dynamical systems or to problems in greater 
dimension, we prefer here to examine a quite different situation which 
in fact can be discussed with a similar procedure. Let us consider a problem 
with Hamiltonian (1), where now the potential is such that
$$V(x)<0 \q , \q \forall x\in\R \qq {\rm with} \qq 
\lim_{x\to\pm\infty}V(x)=0 \en(23)$$
and there is only a stationary point $x_0$ (a minimum) for $V(x)$:
$${\d V(x_0)\over{\d x}}=0 \en(23')$$
There is then a (heteroclinic) orbit approaching $-\infty$ and $+\infty$:
$$\chi=\chi(t) \qq{\rm with}\qq \lim_{t\to\mp\infty}\chi(t)=\mp\infty
\en(24)$$
A well known example is given by the classical Sitnikov restricted 
$3-$body problem in celestial mechanics, which -- at the limit of 
zero eccentricity -- is described by the potential~[4] 
$$V(x)={-1\over{\sqrt{x^2+{1\over 4}}}}\en(25)$$
More in general, we can consider a potential such that
$$V(x)\sim {1\over{|x|^\mu}}\q {\rm for}\q |x|\to\infty\qq{\rm with}\q 
\mu>0,\ {\rm real} \en(26)$$
Clearly, these problems are, {\it per se}, \nhy , the equilibrium point 
being at the infinity; actually, Moser [4] examined the above problem (25), 
in the presence of the periodic perturbation produced by nonzero eccentricity, 
using the McGehee [16] singular coordinate transformation defined by 
$x=2/y^2$ and $\d t=(4/y^3)\ \d s$, which in fact
transforms the problem into a hyperbolic one near the point $y=0$, and he
was able to prove the existence of Smale horseshoe dynamics (see also 
[10,11]). 

Our above procedure can be equally well used in this new case, i.e.
for problems with potential of the form (23,27): indeed, writing the  problem 
as in (8), we can again introduce the variation equation (10-11). 
In this case, one finds that two independent solutions of the homogeneous 
part of (10) have the following  behaviour,  for $|t|\to\infty$
$$\eqalign{ \.\chi(t)\sim &\ |t|^{-\mu/(2+\mu)}\to 0  \cr
\psi(t)\sim &\ |t|^{(2+2\mu)/(2+\mu)}\to\infty  } \en (27)$$
This allows us to repeat exactly the same arguments as in 
part {\bf 1}. In particular, assuming e.g. a non-dissipative periodic 
perturbation of the form, for large $|x|$,
$$g(\ep,x,\.x,t)\sim\ \ep x^{-n}\sin\omega t\en(28)$$
it is easy to show that the existence of out/in sets is granted if
$$n>2+\mu\en(29)$$
and to show also that the Melnikov function (17) possesses infinitely 
many transversal zeroes. Notice that the Sitnikov $3-$body problem 
above mentioned falls into this situation, indeed -- at the first 
order in the eccentricity $\ep$ -- the perturbation is given by [4]
$$g=\ep{-3x\over{4\Big(x^2+{1\over 4}\Big)^{5/2}}}\cos t $$

As already remarked in part {\bf 1}, the standard Birkhoff-Smale theorem 
cannot be directly used for this \nhy\ case. Here, however, at least for 
the case of Sitnikov problem, we can refer to 
the arguments used in [10] (see especially the Appendix of [10], and also 
[13]), to obtain an equivalence to a ``\nhy\ horseshoe'', 
where contracting and expanding actions are not exponential but 
``polynomial'' in time. Alternatively, in the general case, one may possibly 
resort to the method of ``blowing-up'', devised to 
investigate the properties of \nhy\ singularities by means of suitable 
changes of coordinates [17,18], but presumably a full and general 
treatment of \nhy\ horseshoes is still open.
\bs\bs\pn
{\bf Acknowledgments.} 
\pn
We are grateful to prof.s D. Bambusi, U. Bessi, L. Galgani, and 
J. Mallet-Paret for useful suggestions and bibliografical indications.

\bs\bs

\Ref
[1] V.K. Melnikov, {\it Trans. Moscow Math. Soc.} {\bf 12} (1963) 1

[2]  S. Smale, {\it Bull. Amer. Math. Soc.} {\bf 73} (1967) 747

[3]  Z. Nitecki, Differentiable dynamics (MIT Press, Cambridge, Mass. 1971)

[4]  J. Moser, Stable and random motions in dynamical systems (Princeton 
Univ. Press, Princeton 1973)

[5] S.-N. Chow, J.K. Hale and J. Mallet-Paret, {\it J. Diff. Eq.} {\bf 37} 
(1980) 351

[6]  J. Guckenheimer and P.J. Holmes, Nonlinear oscillations, dynamical 
systems and bifurcations of vector fields (Springer, Berlin 1983)

[7] S. Wiggins, Global bifurcations and chaos (Springer, Berlin 1989)

[8]  M. Santoprete, Thesis, Dept. of Physics, Univ. of Pisa

[9]  J. Casasayas, E. Fontich and  A. Nunes, {\it Nonlinear Diff. Eq. 
Appl.} {\bf 4} (1997) 201

[10] H. Dankowicz and P. Holmes, {\it J. Diff. Eq.} {\bf 116} (1995) 468

[11] C. Robinson, {\it Contemp. Math.} {\bf 198} (1996) 45

[12] J. Casasayas, E. Fontich and  A. Nunes, {\it Nonlinearity} {\bf 5} 
(1992) 1193

[13] Xiao-Biao Lin, {\it Dynamics Reported} {\bf 5} (1996) 99

[14] E.A. Coddington  and  N. Levinson, Theory of ordinary differential 
equations (McGraw-Hill, New York 1955)

[15] V. Rayskin, preprint 1998, Dept. of Math., Texas Univ., Austin.

[16] R. McGehee, {\it  J. Diff. Eq.} {\bf 14} (1973) 70

[17] F. Dumortier, {\it J. Diff. Eq.} {\bf 23} (1977) 53

[18] M. Brunella and M. Miari, {\it J. Diff. Eq.} {\bf 85} (1990) 338

\bye